\documentclass[prd,preprint,superscriptaddress]{revtex4}
\usepackage{revsymb}
\usepackage{amsmath}
\usepackage{graphicx}
\usepackage{bm}
\usepackage{subfigure}
\newcommand{\be}{\begin{equation}}
\newcommand{\ee}{\end{equation}}
\newcommand{\ben}{\begin{eqnarray}}
\newcommand{\een}{\end{eqnarray}}

\begin{document}
\title{Cosmological dynamics with non-linear interactions}

\author{Fabiola Ar\'{e}valo}
\email{fabiola.arevalo@gmail.com}
\affiliation{Departamento de F\'{\i}sica,
Facultad de Ciencias F\'{\i}sicas y Matem\'{a}ticas\\
Esteban Iturra s/n, Barrio Universitario
Universidad de Concepci\'{o}n
Concepci\'{o}n, Chile\\
and \\
Departamento de Ciencias F\'{\i}sicas, Universidad de La Frontera, Temuco, Casilla 54-D, Chile
}

\author{Anna Paula Bacalhau}
\email{annaprb@yahoo.com.br}
\affiliation{Universidade Federal do Esp\'{\i}rito Santo,
Departamento
de F\'{\i}sica\\
Av. Fernando Ferrari, 514, Campus de Goiabeiras, CEP 29075-910,
Vit\'oria, Esp\'{\i}rito Santo, Brazil\\
and\\
Instituto de Cosmologia Relatividade e Astrof\'{\i}sica ICRA - CBPF, Rua Xavier Sigaud, 150, Urca, 22290-180, Rio de Janeiro, Brazil}

\author{Winfried Zimdahl}
\email{winfried.zimdahl@pq.cnpq.br}
\affiliation{Universidade Federal do Esp\'{\i}rito Santo,
Departamento
de F\'{\i}sica\\
Av. Fernando Ferrari, 514, Campus de Goiabeiras, CEP 29075-910,
Vit\'oria, Esp\'{\i}rito Santo, Brazil}
\date{\today}

\begin{abstract}
A non-gravitational, non-linear interaction between dark matter and dark energy may result in a future evolution of the Universe which differs from that of the standard $\Lambda$CDM model.
In particular, the ratio of the energy densities of dark matter and dark energy may approach a stable finite value. For a special case we find a corresponding analytic solution for the interacting two-component dynamics which is consistent with the supernova type Ia (SNIa) data from the Union2 set.
For a broader class of interactions without analytic solutions, a dynamical system analysis classifies stationary points with emphasis on their potential relevance for the coincidence problem. Asymptotically stationary solutions of this kind require a phantom-type ``bare" equation of state of the dark energy which, however, does not lead to a big-rip singularity.
\end{abstract}

\maketitle

\section{Introduction}

Investigating the properties of the cosmological dark sector has become one of the major activities in physics since the detection of the accelerated expansion of the Universe more than a decade ago \cite{SNIa}.
According to the most accepted interpretation, based on Einstein's General Relativity, our Universe is dynamically dominated by two so far unknown components, dark matter (DM) and dark energy (DE). The latter contributes roughly 72\% to the total energy budget, the former about 23\%. Only about 5\% are in the form of conventional, baryonic matter. DE, a substance equipped with a sufficiently large negative pressure accounts for the accelerated expansion, DM is needed for successful cosmic structure formation.
Alternative and complementary support for this interpretation comes from the anisotropy spectrum of the cosmic microwave background radiation \cite{cmb}, from large-scale-structure data \cite{lss}, from the integrated Sachs--Wolfe effect \cite{isw}, from baryonic acoustic
oscillations \cite{eisenstein} and from gravitational lensing \cite{weakl}.
The preferred model is the $\Lambda$CDM model which also plays the role of a reference model for alternative approaches to the DE problem.
Because of the cosmological constant problem in its different facets and the coincidence problem, i.e. the question, why the ratio of the energy densities of DM and DE is of the order of unity at the present epoch, a host of alternative models has been developed in which the cosmological term is dynamized. Overviews of the situation can be found, e.g., in \cite{rev,pad,dumarev}. While in the $\Lambda$CDM model and in most of the alternative approaches DM and DE are considered as independent components of the cosmic medium, there exists a line of research that admits a coupling between both dark components.
Such coupling does not contradict the overall energy-momentum conservation.
Interacting models of this type give rise to a richer dynamics and are particularly useful to address the coincidence problem.
Models, applicable to an interaction between DE and DM
were introduced by Wetterich \cite{wetterich}. Meanwhile there exists a still growing body of
literature on the subject -see, e.g., \cite{interacting,dascorasaniti,verde,
amendquartin,wuyu,
quartin,german,chengong,maartensdyn,boehmer,mangano,baldi,hewang,magongchen,
lip,zhang,forte,mazumdar,
abdalla} and
references therein.
A general problem here is the choice of the interaction term. Since neither the physical nature of DE nor the physical nature of DM are known, there does not exist a sound microphysical motivation for a specific interaction either. Therefore, approaches to interacting dark energy are largely phenomenological.
Even though, it has been argued that, being unaware of the possibility of an interaction between both dark components, may result in a misled interpretation of observational data \cite{dascorasaniti}.
Most of the interactions studied so far are linear in the sense that the interaction term in the individual energy balances of the components is proportional either to the DM density or to the DE density or to a linear combination of both densities (for a recent analysis see, e.g., \cite{verde}).
Since systems with interactions admit analytical solutions only in special cases, several authors resorted to a dynamical system analysis (see, e.g., \cite{elliswain,coley,lynch}   for the general background) to obtain a qualitative picture
of the long-time behavior of the cosmological dynamics \cite{amendquartin,wuyu,quartin,german,chengong,maartensdyn,boehmer}.
Occasionally, non-linear couplings were studied as well \cite{mangano,baldi}.
A product coupling, i.e., an interaction proportional to the product of DM density and DE density was shown to be favored observationally over linear choices in \cite{hewang}.
Also from a physical point of view such type of coupling seems preferred. An interaction between two components should depend on the product of the abundances of the individual components, as, e.g., in chemical reactions.
A dynamical system analysis for a specific non-linear interaction was performed  and contrasted with observational results in \cite{magongchen,lip}.

 It is the purpose of this paper to investigate the long-time behavior for a simple two-component model with a number of non-linear interactions on the basis of Einstein's theory.  We show that for a particular case there exists an analytic solution with a non-vanishing, positive limit for the ratio of the energy densities of DM and DE. In this case, the equation-of-state (EoS) parameter $w$ of the DE is necessarily of the phantom type, i.e., $w < -1$. Similar stationary points for a larger class of interactions are found with the help of a dynamical system analysis. We classify these points  according to their possible relevance for the coincidence problem. The existence of these points requires $w < -1$ as well.
 However, while non-interacting models with constant $w < -1$ necessarily approach a singularity after a finite time \cite{caldwell,caldwell2}, the (non-linear) interaction quite generally prevents the cosmic evolution from a final big-rip.
Moreover, independent of the details of the interaction, a positive total energy density in the critical points necessarily implies an energy transfer from DE to DM. We discuss attractors, stable focuses and centers as potential final states of the cosmic dynamics. The center-solution discussed in \cite{lip} is recovered as a particular case of our work.

The paper is organized as follows. In section \ref{general} we present the basic relations of the general interacting two-component model. We clarify that the relevant critical points require a phantom-type EoS parameter and
we show that a positive critical density is only compatible with a production of DM at the expense of DE but not with a transfer in the opposite direction.  A broad class of non-linear interactions is introduced in section \ref{class}, where we also test a particular analytic solution against SNIa data from the Union2 set \cite{Amanullah}.
In section \ref{dynamical} we perform a dynamical system analysis. A  discussion of how the  long-time limit might fit into a viable cosmological scenario is given in section \ref{scenario}. Section \ref{discussion} summarizes the results of the paper.

\section{General analysis}
\label{general}
The dynamics of our present Universe is assumed to be dominated by DE and DM.
The relevant field equations for the spatially flat, homogeneous and isotropic case then
are the Friedmann  equation
\begin{equation}
3\,H^{2} = 8\,\pi\,G\,\left(\rho_{m} + \rho_{x}\right)\ ,
\label{friedmann}
\end{equation}
and
\begin{equation}
\dot{H}\, = - 4\,\pi\,G\,\left(\rho_{m} + \rho_{x} + p_{x}\right)
\ .\label{dotH}
\end{equation}
Here, $\rho_{m}$ is the energy density of pressureless DM and $\rho_{x}$ is
the density of a DE component with a pressure $p_{x}$. Throughout this paper we use units with $c=1$.
We assume
that both components do not conserve separately but interact with
each other  such that the balance equations take the
forms
\begin{equation}
\dot{\rho}_{m} + 3H \rho_{m} = Q\,  \label{dotrhom}
\end{equation}
and
\begin{equation}
\dot{\rho}_{x} + 3H (1+w)\rho_{x} = - Q\, , \label{dotrhox}
\end{equation}
where $w\equiv \frac{p_{x}}{\rho_{x}}$ is the EoS parameter of the dark energy.
The sum of (\ref{dotrhom}) and (\ref{dotrhox}) results in the total energy conservation equation
\begin{equation}
\dot{\rho} + 3H \left(\rho + p\right) = 0\, , \label{dotrhot}
\end{equation}
where the total pressure
equals the dark energy pressure, $p = p_{x}$.
To address the coincidence problem it is convenient to introduce the ratio $r \equiv  \frac{\rho_{m}}{\rho_{x}}$
of the energy densities, which is characterized by
the dynamics
\begin{equation}
\dot{r} = r\left[\frac{\dot{\rho}_{m}}{\rho_{m}} - \frac{\dot{\rho}_{x}}{\rho_{x}}\right]\ .
\label{}
\end{equation}
Furthermore, it is also convenient to introduce an effective pressure quantity $\Pi$ by $Q = - 3H \Pi$ and to replace the derivatives with respect to the cosmic time by derivatives with respect to $\ln a^{3}$, denoted by a prime, i.e.
$\dot{\rho} \equiv \rho^{\prime} 3H$. Then the dynamics of the two-component system is given by
\begin{equation}
\frac{\rho_{m}^{\prime}}{\rho_{m}} = - 1 - \frac{\Pi}{\rho_{m}}\ , \qquad
\frac{\rho_{x}^{\prime}}{\rho_{x}} = - \left(1 + w\right) + \frac{\Pi}{\rho_{x}}\ ,
\label{}
\end{equation}
or, alternatively, by
\begin{equation}
\rho^{\prime} = - \left(1 + \frac{w}{1+ r}\right)\rho\
\label{rhopr}
\end{equation}
and
\begin{equation}
r^{\prime} = r \left[w - \frac{\left(1 + r\right)^{2}}{r \rho}\,\Pi\right]
\, .
\label{rpr}
\end{equation}
All the details of the interaction are encoded in the function $\Pi$.
In the interaction-free limit $\Pi = 0$, the stationary point $r_{s} = 0$ together with $w=-1$ corresponds to the de-Sitter space as the long-time limit of the $\Lambda$CDM model.

The relevant critical points of Eq.~(\ref{rhopr}) are given by
\begin{equation}
r_{c} = - 1 - w\ ,
\label{rc}
\end{equation}
where the subscript $c$ denotes the critical point.
Consequently, for positive values of $r$, the existence of a critical point requires an EoS parameter
$w< -1$, i.e., DE of phantom type. This conclusion does not depend on the interaction. A non-zero stationary value for the ratio $r$ can be interpreted as an alleviation of the coincidence problem.
The condition $r^{\prime} =0$ together with (\ref{rpr}) and (\ref{rc}) provides us with
\begin{equation}
\rho_{c} = - \frac{w}{1+w}\Pi_{c}\ .
\label{rhoc}
\end{equation}
In general, $\Pi_{c} = \Pi_{c}(\rho_{c}, r_{c})$. Therefore  (\ref{rhoc}) is not an explicit relation
for $\rho_{c}$. Moreover, $\rho_{c}$ remains undetermined for a linear dependence of $\Pi$ on $\rho$.  As will be shown in section \ref{dynamical} below in more detail, this case is degenerate and does not admit a dynamical system analysis. On the other hand, for $\Pi \propto \rho$ equation (\ref{rpr}) decouples which will allow us to obtain analytic solutions of the system (\ref{rhopr}) and (\ref{rpr}).

Since $w< -1$, a positive stationary energy density $\rho_{c}$ in (\ref{rhoc}) requires $\Pi_{c} <0$, equivalent to $Q_{c}>0$. Independent of the specific interaction (excluding only a linear dependence $\Pi \propto \rho$), the existence of the critical points $r_{c}$ and $\rho_{c}$ requires a transfer from DE to DM. We disregard here the critical points with $r_{c} = \rho_{c} = 0$ and the unphysical $r_{c} = -1$ ($\rho_c=0$).
We emphasize that as long as $\Pi \propto \rho$ is excluded, the results for the critical points so far do not depend on the structure of $\Pi$.
In the following section we shall consider a specific class of interactions.

\section{A class of non-linear interactions}
\label{class}

\subsection{Structure of the coupling term}

As already mentioned, lacking a microphysically motivated interaction, one has to resort to phenomenological models. Notice that the knowledge of a reliable microphysical interaction
would be equivalent to already knowing the physical nature of DE.
Our principal interest in this paper are non-linear interactions, i.e. interactions for which
the effective pressure $\Pi$ in general is a non-linear function of the energy-densities of the components and/or the total energy density. Motivated by the structure
\begin{equation}
\rho_{m} = \frac{r}{1+r}\rho\quad \mathrm{and}\quad \rho_{x} = \frac{1}{1+r}\rho
\
\label{rmrx}
\end{equation}
of the components, we consider the ansatz
\begin{equation}
\Pi = - \gamma \rho^{m}r^{n}\left(1+r\right)^{s} = - \gamma\rho^{m+s}\rho_{m}^{n}\rho_{x}^{s-n}
\ ,
\label{ansatz}
\end{equation}
where $\gamma$ is a positive coupling constant with a dimension of $\rho^{1-m}$. The powers $m$, $n$ and $s$ specify the interaction. For fixed values $m$, $n$ and $s$ the only  free parameter is  $\gamma$.
A linear dependence of $\Pi$ on $\rho$ corresponds to $m=1$. While this case is not accessible to a dynamical system analysis (see the comments following eq.~(\ref{rhoc}) and section \ref{dynamical} below) it is perfectly admissible in the general dynamics (\ref{rhopr}) and (\ref{rpr}).
Moreover, as already mentioned, it is exactly this case which will provide us with analytic solutions of the system (\ref{rhopr}) and (\ref{rpr}).

The effective interaction pressure $\Pi$ is proportional to powers of products
of the densities of the components for the special cases $s = - m$, but we shall admit $s$ and $m$ to be arbitrary for the moment.
Notice that every power of
the total energy density $\rho$ in the interaction pressure itself corresponds, via Friedmann's equation, to the square of the Hubble parameter ($\rho \propto H^{2}$). This implies that the interaction quantity $Q$ is not necessarily linear in the Hubble rate.
For $s = - m$ the ansatz (\ref{ansatz}) is equivalent to
\begin{equation}\label{Qs=-m}
Q = 3 H \gamma \rho_{x}^{m-n}\rho_{m}^{n} = 3H\gamma \rho_{x}^{m}\,r^{n}\ .
\end{equation}
The ansatz (\ref{ansatz}) contains a large variety of interactions that have been studied in the literature as special cases.
This comprises, e.g., the models in \cite{wuyu,german,quartin,zhang,forte,mazumdar,abdalla}.

As mentioned before, in most of the interacting models in the literature, the interactions $\Pi$ are assumed to be linear in either $\rho_{m}$ or $\rho_{x}$. The corresponding source terms are $Q = 3\gamma H\rho_{m}$ or $Q = 3\gamma H\rho_{x}$, respectively, or a combination of both (see, e.g.,~\cite{wuyu,german,quartin,abdalla}). In our setting, the case $Q = 3\gamma H\rho_{m}$ is recovered for $(m,n,s)=(1,1,-1)$, while the combination $(m,n,s)=(1,0,-1)$ reproduces $Q = 3\gamma H\rho_{x}$.

In the following subsection we consider particular combinations of the parameters $(m,n,s)$ which give rise to analytically solvable models with non-linear interaction terms.

\subsection{Analytically solvable models}

\subsubsection{The case $Q = 3H\gamma\frac{\rho_{m}\rho_{x}}{\rho}$}

This example for an analytically solvable non-linear interaction model, covered by the ansatz (\ref{ansatz}), follows for $(m,n,s)=(1,1,-2)$.
In such a case, equation (\ref{rpr}) reduces to
\begin{equation}
r^{\prime} = r \left[w + \gamma\right]
\, ,
\label{rpran}
\end{equation}
which results in a power-law solution
\begin{equation}
r = r_{0} a^{3\left(w + \gamma\right)}
\, .
\label{rsan}
\end{equation}
The ratio $r$ decreases for $w+\gamma < 0$.
Introducing  (\ref{rsan}) into (\ref{rhopr}), we find the energy density
\begin{equation}
\rho = \rho_{0} a^{-3\left(1+w\right)}\left[\frac{1+r_{0} a^{3\left(w + \gamma\right)}}{1+r_{0}}\right]^{\frac{w}{w+\gamma}}
\, .
\label{rhosolan}
\end{equation}
This case coincides with the interacting model studied in \cite{Scaling,Scaling1,Somasri}, relying on an ansatz $r=r_{0}a^{-\xi}$ for the energy-density ratio. This ansatz was proposed in \cite{Dalal} in order to address the coincidence problem. The correspondence is $\gamma = - \left(w + \frac{\xi}{3}\right)$.
For $a\ll 1$ we have $\rho \propto a^{-3}$, for $a\gg 1$ the behavior is
$\rho \propto a^{-3\left(1+w\right)}$. The $\Lambda$CDM model is recovered for $\xi = 3$ and $w=-1$, equivalent to $\gamma = 0$.
The densities of the components are
\begin{equation}
\rho_{m} = \rho_{m0} a^{-3\left(1-\gamma\right)}\left[\frac{1+r_{0} a^{3\left(w + \gamma\right)}}{1+r_{0}}\right]^{-\frac{\gamma}{w+\gamma}}
\,
\label{rhomsolan}
\end{equation}
and
\begin{equation}
\rho_{x} = \rho_{x0} a^{-3\left(1+w\right)}\left[\frac{1+r_{0} a^{3\left(w + \gamma\right)}}{1+r_{0}}\right]^{-\frac{\gamma}{w+\gamma}}
\, ,
\label{rhoxsolan}
\end{equation}
where $\rho_{m0} = \frac{r_{0}}{1+r_{0}}\rho_{0}$ and $\rho_{x0} = \frac{1}{1+r_{0}}\rho_{0}$, respectively.
The non-interacting limit is correctly reproduced for $\gamma = 0$.
For $w+\gamma = 0\ $ the energy-density ratio $r$ is constant and
\begin{equation}
r = r_{0} \quad \Rightarrow\quad \rho = \rho_{0}a^{-3\left(1+\frac{w}{1+ r}\right)} \ .
\label{r0}
\end{equation}
The model based on (\ref{rhosolan}) has been analyzed in some detail in the literature \cite{Somasri,jailson,david}.  It represents a testable alternative to the $\Lambda$CDM model.
Although the latter is largely consistent with observations, the data leave sufficient room for deviations from either $\xi = 3$ or $w=-1$  which would correspond to a non-vanishing interaction, i.e., $\gamma \neq 0$.

\subsubsection{The case $Q = 3H\gamma\frac{\rho_{m}^{2}}{\rho}$}

 This case corresponds to a choice $(m,n,s)=(1,2,-2)$. It has the analytic solutions
\begin{equation}
r = r_{0} \frac{w}{\left(w+\gamma r_{0}\right)a^{-3w} - \gamma r_{0}}
\,
\label{riii}
\end{equation}
and
\begin{equation}
\rho = \rho_{0}a^{-3\left(1 - \frac{w\gamma}{w - \gamma} \right)}
\left[\frac{\left(w + \gamma r_{0}\right)a^{-3w} + r_{0}\left(w-\gamma\right)}{w\left(1+r_{0}\right)}\right]^{\frac{w}{w-\gamma}}
\, .
\label{rhoiii}
\end{equation}
The high-redshift limit of (\ref{riii}) is
\begin{equation}
r \quad \rightarrow\quad \frac{|w|}{\gamma}\qquad\qquad (a\ll 1)
\, .
\label{riii<}
\end{equation}
For $a\ll 1$, i.e. in the past, the ratio $r$ becomes constant. In the opposite limit $a\gg 1$ on the other hand,
we find $r \propto a^{-3}$ as in the $\Lambda$CDM case.
Assuming $w=-1$, the energy density for small values of the scale factor behaves as
\begin{equation}
\rho \propto a^{-\frac{3}{1+\gamma}}\qquad\qquad (a\ll 1)
\, .
\label{rhoiii<}
\end{equation}
The interaction constant modifies the typical $a^{-3}$ behavior at high redshifts.
In the opposite limit $a\gg 1$, the energy density behaves as $\rho \propto a^{-3(1+w)}$ which coincides with the corresponding dependence of the $w$CDM model. In this case, the interaction does not lead to a different future evolution of the Universe.

\subsubsection{The case $Q=3H\gamma\frac{{\rho_x}^2}{\rho}$}

 The third and most interesting analytical solution corresponds to the choice $(m,n, s)=(1,0,-2)$. For $w<0$, i.e. $w=-|w|$, the solutions are
\begin{equation}
r = \left(r_{0} - \frac{\gamma}{|w|}\right)a^{-3|w|} + \frac{\gamma}{|w|}
\,
\label{rii+}
\end{equation}
and
\begin{equation}
\rho = \rho_{0}a^{-3\left(1-\frac{|w|^{2}}{|w|+\gamma}\right)}
\left[\frac{|w| +\gamma + \left(|w| r_{0}-\gamma\right)a^{-3|w|}}{|w|\left(1+r_{0}\right)}\right]^{\frac{|w|}{|w|+\gamma}}
\, .
\label{rhoii2}
\end{equation}
For $\gamma >0$, positivity of both $r$ and $\rho$ is guaranteed for $\gamma < |w|r_0$:
\begin{equation}\label{}
\gamma < |w|r_0 \quad \Rightarrow\quad r > 0 \quad \mathrm{and} \quad \rho > 0\ .
\end{equation}
The ratio (\ref{rii+}) scales as $r \propto a^{-3|w|}$ for $a\ll 1$. For $w=-1$ this coincides with the scaling of its $\Lambda$CDM counterpart. In the far-future limit, however, we have
\begin{equation}
r \quad \rightarrow\quad \frac{\gamma}{|w|}\qquad\qquad (a\gg 1)
\, ,
\label{rii>}
\end{equation}
i.e., the energy-density ratio remains finite, whereas it tends to zero in the $\Lambda$CDM model.
The energy density scales as $a^{-3}$ for $a\ll 1$, i.e., we recover an early matter dominated period. In the limit $a\gg 1$ one has
\begin{equation}
\rho \propto a^{-3\left(1-\frac{|w|^{2}}{|w|+\gamma}\right)}\qquad\qquad (a\gg 1)
\, ,
\label{rhoii>}
\end{equation}
which generally does not correspond to a de Sitter phase.

The solution (\ref{rhoii2}) has the interesting special case $1-\frac{|w|^{2}}{|w|+\gamma} = 0$ in which $\rho$ tends to a constant for $a\gg1$.
Under this condition we have
\begin{equation}\label{g/w}
|w|^{2} = |w| + \gamma\quad \Rightarrow\quad \gamma = |w|\left(|w|-1\right)\quad \Rightarrow\quad
\frac{\gamma}{|w|} = |w|-1 \ .
\end{equation}
The interaction constant $\gamma$ is directly related to the deviation of $w$ from $w=-1$.
Then, the energy density (\ref{rhoii2}) may be written as
\begin{equation}
\rho = \rho_{0}
\left[\frac{|w| +\left[r_{0}-\left(|w| - 1\right)\right]a^{-3|w|}}{1+r_{0}}\right]^{\frac{1}{|w|}}
\, .
\label{rhofin}
\end{equation}
For $r$ we find
\begin{equation}
r = \left[r_{0} - \left(|w|-1\right)\right]a^{-3|w|} + |w|-1
\, .
\label{rfin}
\end{equation}
The limiting values for $a\gg 1$ are
\begin{equation}
\rho_{\infty} = \rho_{0}
\left[\frac{|w|}{1+r_{0}}\right]^{\frac{1}{|w|}}
\,
\label{rhoinf}
\end{equation}
and
\begin{equation}
r_{\infty} =  |w|-1
\, .
\label{rinf}
\end{equation}
The dynamics results in stationary values for $\rho$ and $r$.
Notice that $\rho > \rho_{\infty}$ and $r > r_{\infty}$. Moreover, the limiting value (\ref{rinf}) for
$r$ coincides with the stationary value (\ref{rc}). In particular, we have again
$w<-1$.
To the best of our knowledge, this solution has not been considered before.
The role of the interaction in the limiting cases is as follows.
In the distant past, e.g. at high redshift, we have $r\gg 1$ and $\rho \approx \rho_{m}$.
Then, $\frac{|\Pi|}{\rho} \approx \frac{|\Pi|}{\rho_{m}} \propto r^{-2} \ll 1$, i.e., the interaction
is negligible and a matter dominated phase is correctly reproduced.
On the other hand, in the far future, $r\ll 1$ and $\rho \approx \rho_{x}$ are valid.
Consequently, $\frac{|\Pi|}{\rho} \approx \frac{|\Pi|}{\rho_{x}} \approx \gamma$ and
$\frac{|\Pi|}{\rho_{m}} \approx \frac{\gamma}{r}$. In this limit the interaction is crucial.
In other words, the interaction is switched on during the cosmic evolution.
In the following we check whether the solutions (\ref{rhofin}) and (\ref{rfin})  with the final stationary values (\ref{rhoinf}) and (\ref{rinf}), respectively, are consistent with current SNIa observations.  The crucial quantity is the Hubble rate that corresponds to the energy density (\ref{rhoii2}):
\begin{equation}
H = H_{0}a^{-\frac{3}{2}\left(1-\frac{|w|^{2}}{|w|+\gamma}\right)}
\left[\frac{|w| +\gamma + \left(|w| r_{0}-\gamma\right)a^{-3|w|}}{|w|\left(1+r_{0}\right)}\right]^{\frac{1}{2}\frac{|w|}{|w|+\gamma}}
\, .
\label{Hubble}
\end{equation}
The deceleration parameter changes from $q = \frac{1}{2}$ for $a\ll 1$ to
\begin{equation}\label{q>}
q = \frac{1}{2}\left[1 - \frac{3|w|^{2}}{|w| + \gamma}\right]\qquad \quad (a\gg 1)\ .
\end{equation}
Notice that for the case (\ref{g/w}) the limiting value is $q = -1$, although $r_{\infty}> 0$ according to (\ref{rinf}), i.e., different from the $\Lambda$CDM model there remains a non-vanishing matter fraction.

\begin{figure}[ht]
\center
\subfigure[]{\includegraphics[width=0.3\textwidth]{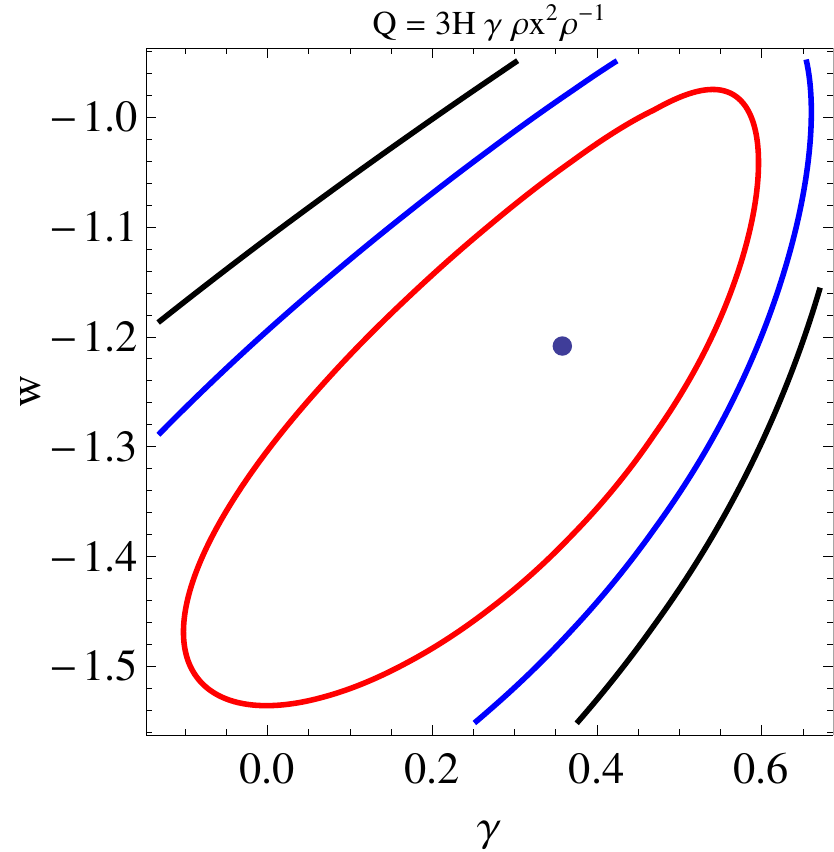} }
\qquad
\subfigure[]{\includegraphics[width=0.3\textwidth]{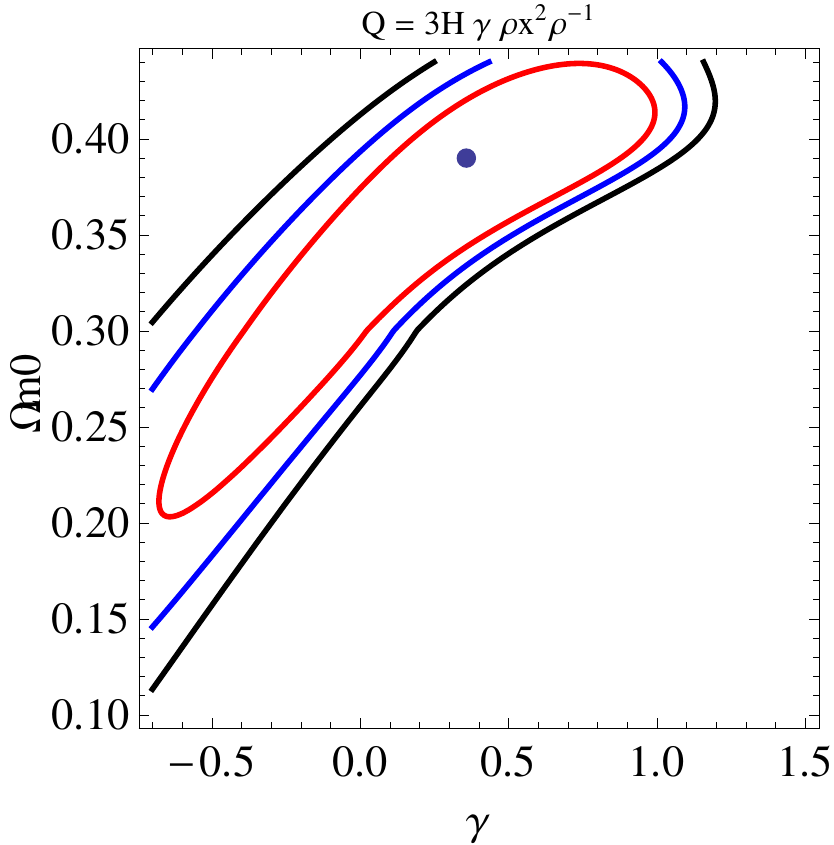}}
\qquad
\subfigure[]{\includegraphics[width=0.3\textwidth]{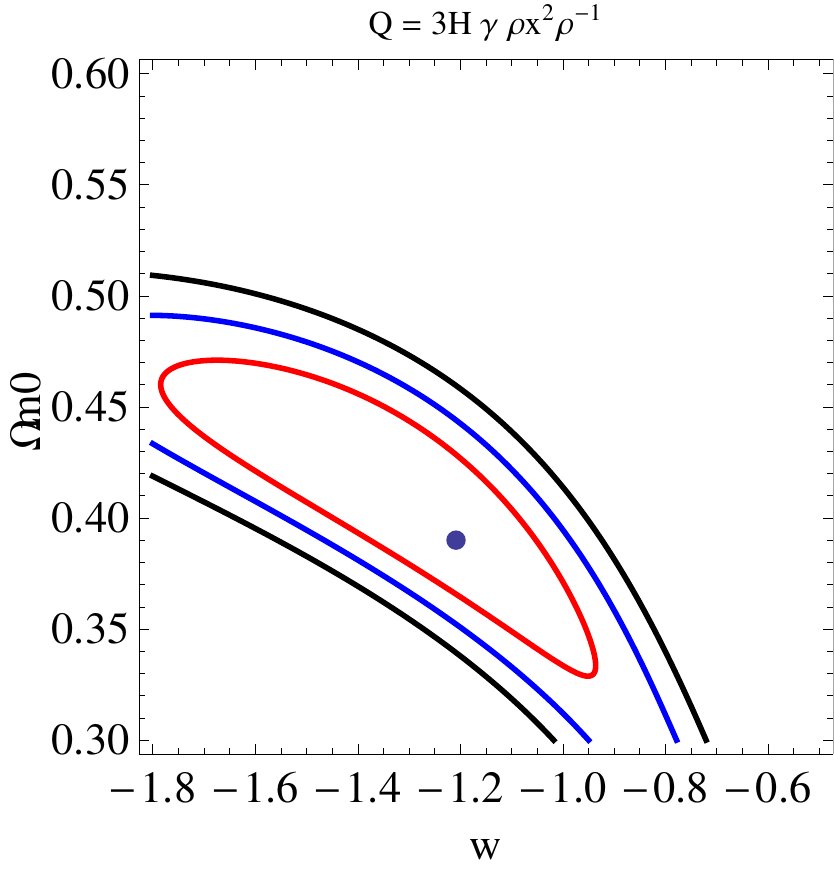}}
\caption{Interaction model with $(m,n,s)=(1,0,-2)$. The contour plots denote the 1$\sigma$, 2$\sigma$ and 3$\sigma$ regions, based
on the Union2 data set.}
\label{modeliicont}
\end{figure}

For our statistical analysis we do not specify beforehand to the solutions (\ref{rhofin}) and (\ref{rfin}). Our aim is to clarify whether the parameter combination (\ref{g/w}) that corresponds to these solutions has observational support.
The free parameters are $|w|$, $\gamma$ and $\Omega_{0} = \frac{r_{0}}{1+r_{0}}$.
We performed a Bayesian statistical analysis on the basis of the SNIa data from the Union2 set \cite{Amanullah}, using the marginalization method for $H_{0}$ developed in \cite{lazkoz}. The results of this analysis are shown in Fig.~\ref{modeliicont}.
The best-fit parameter values are $(\gamma, w, \Omega_{0}, {\chi^2}_{min})= (0.36_{-1.25}^{+0.60},-1.20_{-0.4}^{+0.26}, 0.39_{-0.19}^{+0.08}, 540.863)$.
These results demonstrate, that within the $1\sigma$ region the solutions (\ref{rhofin}) and (\ref{rfin})  with the final stationary values (\ref{rhoinf}) and (\ref{rinf}), respectively, are indeed consistent with the observational data. This may indicate a phenomenological solution of the coincidence problem with the help of non-linear interactions in the dark sector.

For an arbitrary combination of the parameters $m$, $n$ and $s$, analytical solutions of the non-linear system are hardly available.
To get insight into the behavior of the system under more general conditions we shall resort to a dynamical system analysis in the following section.

\section{A dynamical system analysis}
\label{dynamical}

 In this section we consider the dynamics in the vicinity of the critical points with the help of a dynamical system analysis. This analysis is based on the circumstance that, close to the critical points, the (generally unknown) solution of the non-linear system behaves as the solution of the system, linearized around the critical points (Hartmann's theorem and, for purely imaginary eigenvalues, the Center Manifold Theorem (see, e.g., \cite{lynch} and \cite{boehmer}).
Using standard techniques (see, e.g., \cite{coley,lynch}), the general characteristic equation for the critical points is
\begin{equation}
\lambda^{2} + \left[2+w -w\left(1+w\right)\frac{\partial_{r} \Pi}{\Pi}\right]\lambda
+ \left(1+w + w \partial_{\rho} \Pi\right) = 0
\ .
\label{lambda}
\end{equation}
Here $\partial_{r} \Pi$ and $\partial_{\rho} \Pi$ denote the partial derivatives of $\Pi$ with respect to $r$ and to $\rho$, respectively.
Eq.~ (\ref{lambda}) has the solutions
\begin{equation}
\lambda_{\pm} = \frac{1}{2}\left\{\left[w\left(1+w\right)\frac{\partial_{r} \Pi}{\Pi}
- \left(2+w\right)\right]
\pm \sqrt{\left(2+w - w\left(1+w\right)\frac{\partial_{r} \Pi}{\Pi}\right)^{2}
- 4\left(1+w + w\partial_{\rho} \Pi\right)}\right\}
\ ,
\label{lambdasol}
\end{equation}
where we have to require $1+w + w\partial_{\rho} \Pi \neq 0$.
In case these solutions are non-degenerate and real, they describe an attractor for $\lambda_{\pm} < 0$,
an unstable critical point for $\lambda_{\pm} > 0$ and a saddle if $\lambda_{+}$ and $\lambda_{-} $ have different signs \cite{lynch}.
For complex eigenvalues $\lambda_{\pm} = \alpha +i\beta$, it is the sign of $\alpha$ that determines the character of the stationary point. For $\alpha = 0$ the critical point is a center, for $\alpha < 0$ it is a stable focus and for $\alpha > 0$ it is an unstable focus.
With the ansatz (\ref{ansatz}) for the interaction, the eigenvalues (\ref{lambdasol}) are
\begin{equation}
\lambda_{\pm} = -\frac{1}{2}\left\{\left[2+ s + \left(1+n+s\right)w\right]
\mp\sqrt{\left(2+ s + \left(1+n+s\right)w\right)^{2}
+ 4\left(m-1\right)\left(1+w\right)}\right\}
\ .
\label{lambdasol2}
\end{equation}
Since the analysis is valid only for non-zero eigenvalues, all cases with $m=1$, corresponding to a linear dependence of $\Pi$ on $\rho$, are not covered by the classification mentioned before.
  Notice  that both the linear interactions and the analytically solvable non-linear cases of the previous section have $m=1$. For $m=1$ equation (\ref{rpr}) is decoupled from $\rho$ and can be solved separately.

 For $m\neq 1$ the general classification provides us with the following set of critical points:
\begin{itemize}
  \item Attractor for $m>1$ and $s < - \frac{2+\left(1+n\right)w}{1+ w} - 2\sqrt{\frac{1-m}{1+w}}$
  \item Unstable for $m>1$ and $s > - \frac{2+\left(1+n\right)w}{1+ w} + 2\sqrt{\frac{1-m}{1+w}}$
  \item Saddle for $m< 1$, for all $n$ and $s$
  \item Center for $m>1$ and $2+ s + \left(1+n+s\right)w = 0$  for $n>1$
  \item Stable focus for $m>1$ and $2+ s + \left(1+n+s\right)w > 0$
  \item Unstable focus for $m>1$ and $2+ s + \left(1+n+s\right)w < 0$
\end{itemize}

For the critical density we find ($m\neq 1$)
\begin{equation}
\rho_{c} = \left[\gamma |w|^{s+1}\left(|w| - 1\right)^{n-1}\right]^{\frac{1}{1-m}}
\ .
\label{rocrw}
\end{equation}
The corresponding values for the components are
\begin{equation}
\rho_{mc} = \left[\gamma |w|^{s+m}\left(|w| - 1\right)^{n-m}\right]^{\frac{1}{1-m}}
\
\label{romcrw}
\end{equation}
and
\begin{equation}
\rho_{xc} = \left[\gamma |w|^{s+m}\left(|w| - 1\right)^{n-1}\right]^{\frac{1}{1-m}}
\ ,
\label{roxcrw}
\end{equation}
consistent with $r_{c} = \frac{\rho_{mc}}{\rho_{xc}} = |w|-1$ in (\ref{rc}). Consequently, the fixed points $(\rho_c,r_c)$ are given by (\ref{rocrw}) and (\ref{rc}).
For $m>1$ the critical densities are proportional to a negative power of the interaction constant, i.e., the diverge in the limit $\gamma \rightarrow 0$. This limit corresponds to the big-rip singularity of dark energy models with constant EoS parameters $w<-1$ \cite{caldwell,caldwell2}. In other words, any non-vanishing interaction of the type considered here is big-rip avoiding.

\begin{figure}[ht]
\center
\subfigure[]{\includegraphics[width=0.3\textwidth]{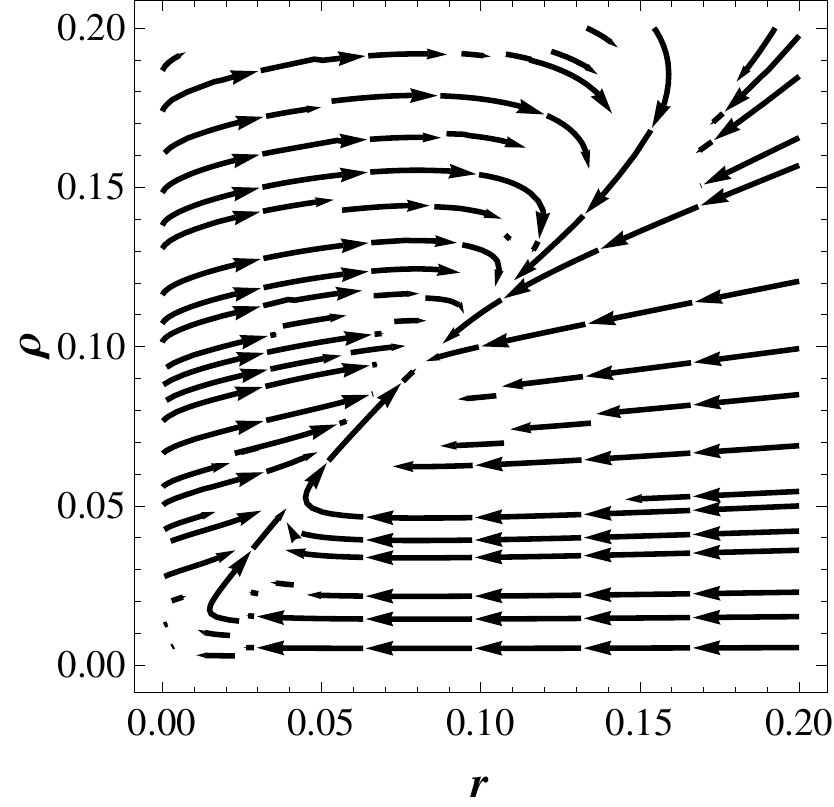} }
\qquad
\subfigure[]{\includegraphics[width=0.3\textwidth]{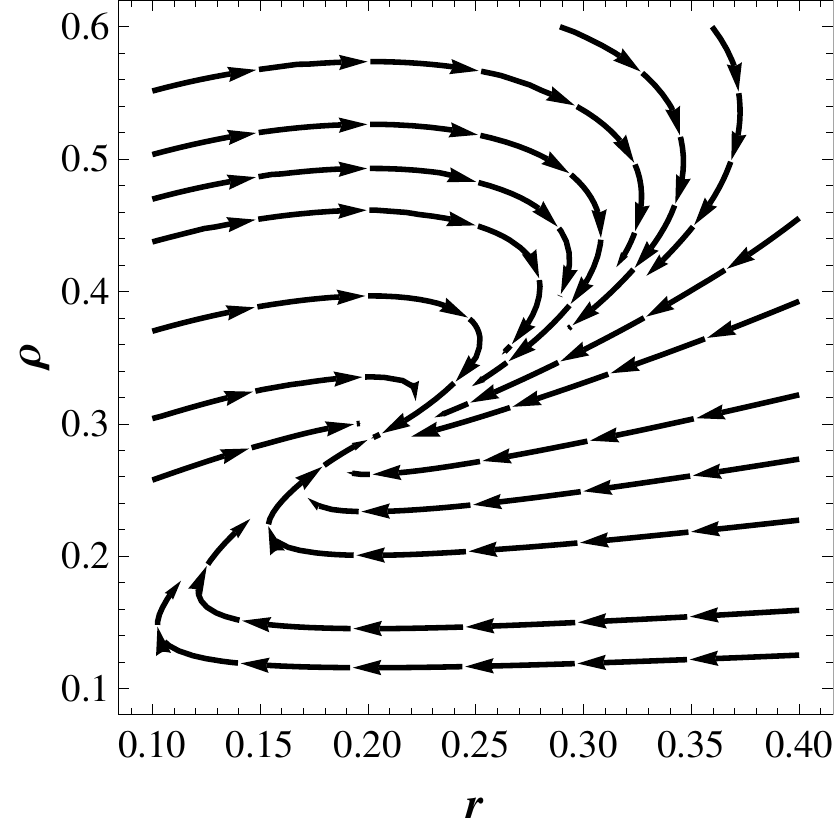}}
\qquad
\subfigure[]{\includegraphics[width=0.3\textwidth]{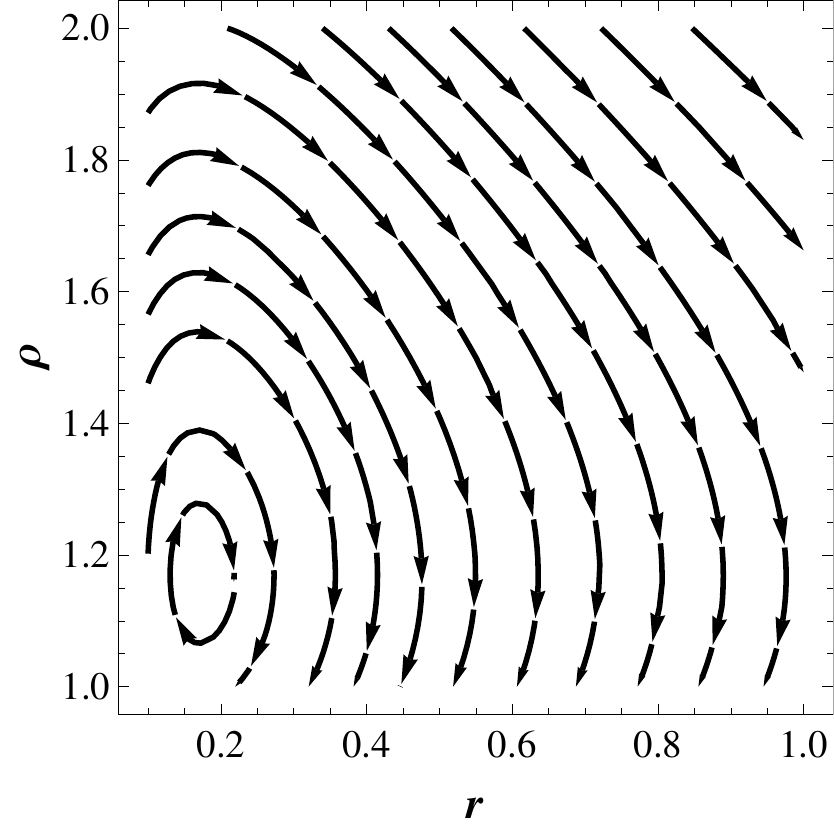}}
\caption{Phase portraits: in (a) the interaction $Q= 3 \gamma H \frac{\rho_m \rho_x}{ \sqrt{\rho_m+ \rho_x}}$ corresponds to an attractor for $w=-1.08$ and $\gamma=1$. The same interaction results in a stable focus for $w=-1.2$ and $\gamma=1$ in (b). In (c) the interaction $ Q= 3 \gamma H \rho_m \rho_x$ describes a center with $w=-1.17$ and $\gamma=1$.}
\label{phaseportraits}
\end{figure}

In tables \ref{attractor}, \ref{sfocus} and \ref{center} we present potentially interesting examples for different critical points, together with the allowed ranges for the EoS parameters. Fig.~(\ref{phaseportraits}) shows the corresponding phase-space portraits.

For a physical interpretation it is instructive to check the analytic solutions of the linearized system that underly (\ref{lambdasol}) and (\ref{lambdasol2}) explicitly for special cases.

(i) An attractor with $(m,n,s)=(2,0,-2)$.
This corresponds to
$\Pi = - \gamma\rho^{2}\left(1+r\right)^{-2}$.
The general expression (\ref{rocrw}) for the critical density reduces to
\begin{equation}
\rho_{c} = \frac{|w|\left(|w| - 1\right)}{\gamma} \ .
\label{rhocatt }
\end{equation}
Eq.~(\ref{rpr}) simplifies to
\begin{equation}
r' = r w + \gamma\rho\ .
\label{dratt}
\end{equation}
Let us introduce quantities $f$ and $g$ which describe small deviations from the critical point:
\begin{equation}
\rho = \rho_{c} + f\ , \qquad r = r_{c} + g\ .
\label{fg}
\end{equation}
While in zeroth order $r_{c}w + \gamma\rho_{c} = 0$ is valid,
we have from (\ref{dratt}) and (\ref{rhopr}), up to linear order in $f$ and $g$,
\begin{equation}
g' = g w + \gamma f\ ,\qquad \mathrm{and} \qquad f' = - \frac{g}{\gamma}\left(|w| - 1\right) \ .
\label{gpratt}
\end{equation}
Eliminating $f$ yields
\begin{equation}
g'' + |w| g'  + \left(|w| - 1\right)\,g = 0\ .
\label{gprprat}
\end{equation}
This second-order equation with constant coefficients has the solution $g = g_{0}\exp{\left[\lambda x\right]}$,
where $g_{0}$ is some initial value and $\lambda$ is determined by $\lambda^{2} + |w|\lambda + \left(|w| - 1\right) = 0$, i.e., $\lambda_{1,2} = - \frac{|w|}{2} \pm \sqrt{\frac{|w|^{2}}{4} - \left(|w| - 1\right)}$.
Both solution are negative and, consistently, represent a special case of (\ref{lambdasol2}). Consequently, $g$ decays exponentially with $x$. Recalling that $x = \ln a^{3}$, we have
\begin{equation}
g = g_{0}\,a^{3\lambda}\qquad \mathrm{and} \qquad f = \frac{|w| + \lambda}{\gamma}g_{0}a^{3\lambda}\ .
\label{ga}
\end{equation}
Since $\lambda$ is negative, $g$ and $f$ decay with a power of the scale factor.
It is the scale-factor dependence of (\ref{ga}) that is behind the curves in Fig.~\ref{phaseportraits}(a).

The expression $\frac{w}{1+r}$ in (\ref{rhopr}) represents the total effective equation of state of the cosmic medium. Close to the critical value $\frac{w}{1+r_{c}}=-1$ it is given by
\begin{equation}
\frac{w}{1+r} = - 1 + \frac{g}{|w|} = - 1 + \frac{g_{0}a^{3\lambda}}{|w|} \ .
\label{weffg}
\end{equation}

The deceleration parameter is related to $\frac{w}{1+r}$ by
\begin{equation}
q = - 1 - \frac{\dot{H}}{H^{2}} \quad \Rightarrow\quad q = \frac{1}{2}\left(1 + 3 \frac{w}{1+r}\right)
\ .
\label{q}
\end{equation}
In the critical point itself $q_{c} = -1$ is valid. In the vicinity of the critical point we have
\begin{equation}
q = - 1 + \frac{3}{2}\frac{g}{|w|}
\
\label{qosc}
\end{equation}
with $g$ from (\ref{ga}). Both the total equation of state $\frac{w}{1+r}$ and the deceleration parameter $q$ approach
$-1$ with the power $3\lambda$ of the scale factor.

\begin{table}[!t]
\begin{center}
\begin{tabular}{|c|c|c|c|c|c|}
  \hline
  $m$ & $n$ & $s$ & $\Pi$ & $Q$ & $w$\\
  \hline
  \hline
  $\frac{3}{2}$ &$0$&$-1$&$ -\gamma\rho^{3/2}\left(1+r\right)^{-1}$  &$3H\gamma \sqrt{\rho}\rho_{x}$ &$-1.5 \leq  w < -1$\\
  \hline
  $\frac{3}{2}$ &$\frac{1}{2}$ &$-\frac{3}{2}$&$ -\gamma\rho^{3/2}r^{1/2}\left(1+r\right)^{-3/2}$  &$3H\gamma \sqrt{\rho_{m}}\rho_{x}$ &$-1.125 \leq  w < -1$\\
  \hline
  $\frac{3}{2}$ &$\frac{1}{2}$&$-1$& $ -\gamma\rho^{3/2}r^{1/2}\left(1+r\right)^{-1}$ &$3H\gamma \sqrt{ \rho \rho_{x}\rho_{m}}$ & $-1.101 \leq  w < -1$\\
  \hline
  $2$ &$\frac{1}{2}$&$-3$&$ -\gamma\rho^{2}r^{1/2}\left(1+r\right)^{-3/2}$ & $3H\gamma \rho_{x}\sqrt{\rho\rho_{m}}$&$-1.0625 \leq w<-1$\\
  \hline
\end{tabular}
\end{center}
\caption{Examples for an attractor as critical point.}  \label{attractor}
\end{table}
\begin{table}[!t]
\begin{center}
\begin{tabular}{|c|c|c|c|c|c|}
  \hline
  $m$ & $n$ & $s$ & $\Pi$ & $Q$ & $w$\\
  \hline
  \hline
  $\frac{3}{2}$ &$0$&$-1$&$ -\gamma\rho^{3/2}\left(1+r\right)^{-1}$  &$3H\gamma \sqrt{\rho}\rho_{x}$ &$w < -1.5$\\
  \hline
  $\frac{3}{2}$ &$\frac{1}{2}$ &$-2$&$ -\gamma\rho^{3/2}r^{1/2}\left(1+r\right)^{-2}$  &$3H\gamma \sqrt{\frac{\rho_{m} \rho_{x}^{3}}{\rho}}$ &$-6.83 \leq  w < -1.17$\\
  \hline
  $\frac{3}{2}$ &$\frac{1}{2}$&$-\frac{3}{2}$& $ -\gamma\rho^{3/2}r^{1/2}\left(1+r\right)^{-3/2}$ &$3H\gamma \sqrt{\rho_{m}}\rho_{x}$ & $w < -1.125 $\\
  \hline
  $\frac{3}{2}$ &$\frac{1}{2}$&$-1$&$ -\gamma\rho^{3/2}r^{1/2}\left(1+r\right)^{-1}$ & $3H\gamma \sqrt{\rho\rho_{m}\rho_{x}}$&$-2 \leq w<-1.101$\\
  \hline
\end{tabular}
\end{center}
\caption{Examples for a stable focus as critical point.}  \label{sfocus}
\end{table}
\begin{table}[!t]
\begin{center}
\begin{tabular}{|c|c|c|c|c|c|}
  \hline
  $m$ & $n$ & $s$ & $\Pi$ & $Q$ & $w$\\
  \hline
  \hline
  $\frac{3}{2}$ &$\frac{3}{2}$&$-\frac{9}{2}$&$ -\gamma\rho^{3/2}r^{1/2}\left(1+r\right)^{-\frac{9}{2}}$  & $3H \gamma \frac{{\rho_{m}}^{\frac{3}{2}} {\rho_{x}}^3}{\rho^3}$ &$w = - 2$\\
  \hline
  $\frac{3}{2}$ &$1$ &$-2$&$ -\gamma\rho^{3/2}r\left(1+r\right)^{-2}$  &$3H\gamma \frac{\rho_{m} \rho_{x}}{\sqrt{\rho}}$ &$w < -1$\\
  \hline
  $2$ &$1$&$-2$& $ -\gamma\rho^{2}r\left(1+r\right)^{-2}$ &$3H\gamma \rho_{m}\rho_{x}$ & $w < -1$\\
  \hline
   $\frac{3}{2}$ & $\frac{3}{2}$& $-5$& $ -\gamma\rho^{3/2}r^{3/2}(1+r)^{-5}$ &$3H\gamma\sqrt{\frac{{\rho_m}^3{\rho_x}^2}{\rho^2}}$ & $w = -1.2$\\
  \hline
\end{tabular}
\end{center}
\caption{Examples for a center as critical point. Notice that the second case here corresponds to the model studied in \cite{lip}.}  \label{center}
\end{table}

(ii) A similar analysis can be made for for the dynamics around a center with $(n,s)=(1,2)$, corresponding to
\begin{equation}
\Pi = - \gamma\rho^{m-2}\rho_{m}\rho_{x}\quad\Leftrightarrow \quad Q = 3H\gamma\rho^{m-2}\rho_{m}\rho_{x}\ .
\label{picenter}
\end{equation}
In this case, the critical density is
\begin{equation}
\rho_{c} = \left(\frac{|w|}{\gamma}\right)^{\frac{1}{m-1}} \ .
\label{rhocm}
\end{equation}
With the ansatz (\ref{fg}) we have, up to linear order,
\begin{equation}
\rho^{m-1} = \rho^{m-1}_{c}\left[1 + \left(m-1\right)\frac{f}{\rho_{c}}\right]\ .
\label{rf}
\end{equation}
The linearized system becomes
\begin{equation}
g' = - \gamma \left(m-1\right)\left(1+w\right)\left(\frac{|w|}{\gamma}\right)^{\frac{m-2}{m-1}}\,f
\qquad \mathrm{and} \qquad
f' = \frac{g}{w}\left(\frac{|w|}{\gamma}\right)^{\frac{1}{m-1}}
\ .
\label{fprg}
\end{equation}
This is equivalent to the second-order equation
$f'' - \left(m-1\right)\left(1+w\right)f = 0$
which, for $m>1$, describes oscillations with a frequency $\omega = \sqrt{\left(1-m\right)\left(1+w\right)}$, i.e.,
$f = f_{0}\cos{\left(\sqrt{\left(1-m\right)\left(1+w\right)}\,x\right)}$
where $x = \ln a^{3}$. For $g$ we have an identical equation with solution
$g = g_{0}\sin{\left(\sqrt{\left(1-m\right)\left(1+w\right)}\,x\right)}$,
i.e., the dynamics is described by ellipses in the $f-g$ plane. This behavior is visualized in Fig.~\ref{phaseportraits}(c)
The change of the total energy density
in (\ref{rhopr}) is given by
\begin{equation}
\frac{\rho'}{\rho} = \frac{f'}{\rho} = - \frac{g}{|w|}
\ .
\label{rhoprfrac}
\end{equation}
For $g>0$, i.e., if the ratio $r$ is larger than the critical value, equivalent to an enlarged matter contribution, we have $\rho' < 0$, i.e., the total energy density decreases since the effective EoS parameter is larger than $-1$.
On the other hand, for $g<0$, there is an excess of dark energy and the dynamics is given by $\rho' > 0$.
The rate of change of $\rho$ is oscillating about the critical point.
The period $\sqrt{\left(1-m\right)\left(1+w\right)}x = 2\pi$ corresponds to $a_{p} = \exp{\left[\frac{2\pi}{3\,\sqrt{\left(1-m\right)\left(1+w\right)}}\right]}$.  For
$m=2$ and $w=-1.1$, e.g., we find $a_{p}\approx 750$ numerically.

In the vicinity of the critical point we have
\begin{equation}
q = - 1 - \frac{3}{2}\frac{g}{w}
\ .
\label{qoscii}
\end{equation}
With an oscillatory solution for $g$ the deceleration parameter oscillates around $q = - 1$. This corresponds to an oscillation of the effective total equation-of-state parameter $\frac{w}{1+r}$ around $-1$ as well. The cosmic medium as a whole oscillates between phantom- and non-phantom behavior, the phantom divide is crossed periodically.

(iii) A mixed case is the stable focus, e.g., the case $(m,n,s)=\left(\frac{3}{2},\frac{1}{2},-1\right)$.
By a similar analysis we find that the perturbations about the critical point behave as damped oscillations with a frequency (with respect
to $x$) of $\frac{1}{2}\sqrt{2\left(|w|-1\right) - \left(1+ \frac{w}{2}\right)^{2}}$ and a damping factor
$a^{-3(1+w/2)/2}$. The critical values $-1$ for the effective equation of state and the deceleration parameter are approached by a corresponding spiralling-in dynamics which is visualized in Fig.~\ref{phaseportraits}(b).

(iv) The case $m=n=s=0$ is an example for a saddle. Here we have $\Pi = $ const. Consistent with the corresponding special case of (\ref{lambdasol2}), only one of the solutions approaches $q=-1$ with a power of the scale factor, the second solution is unstable.

It is interesting to realize that the same interaction may result in different critical points for different ranges of the EoS parameter. The third example in Table \ref{attractor} corresponds  the same expression for $Q$ as the fourth example in Table \ref{sfocus}. For the range $-1.001 \leq w < -1$
the critical point is an attractor while it is a stable focus for $-2 \leq w < -1.101$. A similar situation occurs for the second case  in Table \ref{attractor} and the third case in Table \ref{sfocus}.

\section{An early matter-dominated phase}
\label{scenario}

The focus in this paper is on the late-time behavior of the cosmological dynamics. However, for this behavior to be part of a viable scenario, the dynamics has to be compatible with the present-time observational data and it has to admit an early matter-dominated phase in order to guarantee structure formation.  For the analytic solution (\ref{rhofin}) and (\ref{rfin}) these requirements were satisfied. The situation is less clear for the results of the dynamical system analysis of the last section. To better understand this point, it is useful to write the balances for the components in the form
\begin{equation}\label{rhompr}
\rho_{m}^{\prime} + \rho_{m}\left[1 - \gamma\rho^{m-1}r^{n-1}\left(1+r\right)^{s+1}\right] = 0
\end{equation}
and
\begin{equation}\label{rhoxpr}
\rho_{x}^{\prime} + \rho_{x}\left[1 +w + \gamma\rho^{m-1}r^{n}\left(1+r\right)^{s+1}\right] = 0\ .
\end{equation}
Of particular interest for an acceptable dynamics is a final attractor (or a stable focus) for which both $\rho$ and $r$ approach their final stationary values from $\rho > \rho_{c}$ and $r >r_{c}$ respectively. This implies the requirements $\rho^{\prime} < 0$ and $r^{\prime} < 0$, respectively, during the cosmic evolution. The consequences are
\begin{equation}\label{rhopr<}
\rho^{\prime} < 0 \quad \Rightarrow\quad r > |w| -1 = r_{c}
\end{equation}
and
\begin{equation}\label{rpr<}
r^{\prime} < 0 \qquad \Leftrightarrow\qquad \gamma \rho^{m-1} < |w|r^{1-n}\left(1+r\right)^{-\left(s+2\right)}\ .
\end{equation}
The condition (\ref{rpr<}) puts upper limits on the interaction constant $\gamma$.
For the present epoch we have $r_{0} > |w| -1$ and $\gamma \rho_{0}^{m-1} < |w|r_{0}^{1-n}\left(1+r_{0}\right)^{-\left(s+2\right)}$. The right-hand side of the last inequality is of the order of one. The present energy density $\rho_{0}$ is of the order of the critical density
$\rho_{\mathrm{cr}0} = 1.88 h^{2} 10^{-29}\mathrm{g cm^{-3}}$.
Consequently, $\gamma \lesssim \left(10^{29}\mathrm{g^{-1} cm^{3}}\right)^{m-1}$. (Assuming tentatively the applicability of the latter inequality also for the analytic solution (\ref{rhofin}) which has $m=1$, this condition reduces to $\gamma < 1$, consistent with the result $\gamma \approx 0.36$ of our data analysis.)
It follows from (\ref{rhompr}) and (\ref{rhoxpr}) that for any value of $\gamma$, considerably smaller than this one, the interaction would only provide a very small correction to the dynamics at the present epoch.
On the other hand, $\rho^{\prime} < 0$ implies
\begin{equation}\label{rho0>}
\rho_{0}^{m-1} > \rho_{c}^{m-1} = \frac{|w|^{-\left(s+1\right)}\left(|w| - 1\right)^{1-n}}{\gamma}\ .
\end{equation}
This provides us with
\begin{equation}\label{<<}
|w|^{-\left(s+1\right)}\left(|w| - 1\right)^{1-n} < \gamma \rho_{0}^{m-1} < |w|r_{0}^{1-n}\left(1+r_{0}\right)^{-\left(s+2\right)}\
\end{equation}
and
\begin{equation}\label{w-1}
\left(|w| - 1\right)^{1-n} < \left(|w|^{s+1}\gamma \rho_{0}^{m-1}\right)\ .
\end{equation}
The inequality (\ref{w-1}) relates the interaction strength to deviations from $|w| = 1$ (cf.~Eq.~(\ref{g/w}) for a similar feature of the analytic solutions (\ref{rhofin}) and (\ref{rfin})).
Except for $n=1$, any $\gamma \rho_{0}^{m-1} \ll 1$, equivalent to a small influence
of the interaction on the present cosmological dynamics, requires a value $w$ of the EoS parameter close to $w = -1$. In other words, deviations from $w = -1$ are a measure of the interaction strength. At the same time, a small deviation from  $w = -1$ corresponds to a small, but non-zero limiting value $r_{c} < 1$ of the energy-density ratio. Since a phantom-type EoS parameter close to -1 is preferred by a number of investigations (see, e.g., \cite{montesano} for a recent analysis), a scenario with finite long-time limits $r_{c}$ and $\rho_{c}$  does not seems to contradict current observations.

Let us now consider qualitatively the conditions for the existence of an early matter dominated epoch.
According to (\ref{rhopr}), a matter era with $\rho \propto a^{-3}$ requires $\frac{|w|}{1+r}\ll 1$ for $a\ll 1$. For a constant value of the EoS parameter $|w|$ this is achieved for $r \gg 1$ at $a\ll 1$. Since, according to (\ref{rc}), the far-future limit of $r$ is smaller then unity (for values of $w$ slightly but not substantially smaller than $-1$), one has $r^{\prime} < 0$ as already mentioned, i.e., a decaying ratio of the energy densities during the cosmic evolution.
Assuming accordingly, that at high redshifts $r\gg 1$ is valid,
the balance equation (\ref{rhompr}) can be written
\begin{equation}\label{prms=-m}
\frac{\rho_{m}^{\prime}}{\rho_{m}} \approx - \left[1 - \gamma  r^{m+n+s-1}\rho_{x}^{m-1}\right]\ .
\end{equation}
Then, under the condition $r\gg 1$ and with $m>1$, the interaction term is negligible for  $m+n+s< 1$ in (\ref{prms=-m}), which corresponds to a matter dominated phase. By inspection one realizes that this requirement is met by the first, second and fourth cases of Table \ref{attractor} and by the first, second and third cases of table \ref{sfocus}.
These examples are potential candidates for a scenario which correctly reproduces an early matter dominated period but predicts a future evolution towards an attractor or a stable focus with a finite, non-vanishing value for the ratio of the energy densities. Also the first, second and fourth examples of Table~\ref{center} have $m+n+s< 1$. For the remaining cases which have $m+n+s = 1$ there is no dependence on $r$ and a matter dominated phase requires $\gamma \rho_{x}^{m-1}\ll 1$ for $a\ll 1$.
Although these considerations remain on a heuristic level, its preliminary conclusions do not seem to be inconsistent with a scenario that includes standard structure formation.

\section{Discussion}
\label{discussion}

Models with an interaction between dark matter and dark energy have received considerable attention since they provide a framework to address the cosmic coincidence problem. We have analyzed here  an interacting two-component system of dark matter and dark energy with the total energy density $\rho$ and the ratio of the energy densities of dark matter and dark energy $r = \frac{\rho_{m}}{\rho_{x}}$ as independent variables. We have shown that asymptotically finite, positive critical values $r_{c}$ and $\rho_{c}$ of these variables require a phantom-type equation of state for the dark energy component and an energy transfer from dark energy to dark matter, at least in the critical point. These results hold quite generally and  do not depend on the details of the interaction. We have investigated in some detail interactions of the type $Q = 3 H \gamma \rho^{m+s}\rho_{m}^{n}\rho_{x}^{s-n}$ which generalizes a number of models that have been studied so far in the literature. For various combinations of $m$, $n$ and $s$ the resulting dynamics was shown to be asymptotically different from that of the $\Lambda$CDM model. As a consequence of the interaction, a phantom-type equation of state of the dark energy does not result in a big-rip singularity.
For the particular interaction $Q = 3 H \gamma \rho_{x}^{2}/\rho$ we found an analytic solution that reproduces an early matter-dominated phase and approaches a finite, stationary value for the ratio of the energy densities of DM and DE.
This solution is consistent with the observational data of SNIa from the Union2 data set.
It implies a direct relation between the interaction strength and deviations from $w=-1$.
The interaction for this case is negligible at high redshifts but crucially influences the future dynamics.

Based on a dynamical system analysis for a broader class of interactions, we identified attractors, stable focuses and centers as potential limiting configurations of the cosmological dynamics. For attractors and stable focuses (examples are summarized in Tables \ref{attractor} and \ref{sfocus}, respectively,) the asymptotic value $q_{c} = -1$ of the deceleration parameter is approached by a power of the scale factor.
For centers (see the examples in Table \ref{center}) we find oscillations of the deceleration parameter around this critical point, equivalent to a periodic crossing of the phantom divide for the EoS of the \textit{total} cosmic substratum. All these models have necessarily $m>1$.
A qualitative discussion of the conditions under which this class of models admits the existence of an early matter dominated phase singles out models with $m+n+s<1$.
A more quantitative analysis as well as a study of the perturbation dynamics of non-linearly interacting models will be the subject of future research.

\acknowledgments{Partial support by CNPq and CAPES is gratefully acknowledged. F.A. would like to thank Project MECESUP-FSM0605 for traveling support, the PPGFIS-UFES for their hospitality and CONICYT PHD grant nº21070949 (Chile). The authors would like to thank Roly David Rodriguez Castro for helpful discutions.}

\end{document}